\documentclass[twocolumn,showpacs,pra,superscriptaddress]{revtex4}
\usepackage{dcolumn}
\usepackage{graphicx}
\usepackage{bbm}
\usepackage{amsmath}

\newcommand{\be}{\begin{equation}}
\newcommand{\ee}{\end{equation}}
\newcommand{\beq}{\begin{eqnarray}}
\newcommand{\eeq}{\end{eqnarray}}

\newcommand{\hide}[1]{}

\newcommand{\ket}[1]{\left| #1 \right. \rangle}

\begin{document}

\title{Chirped pulse excitation of two-atom Rydberg states}

\author{Elena Kuznetsova}
\address{Institute of Applied Physics RAS, 46 Ulyanov Street, Nizhny Novgorod, 603950, Russia}
\address{Russian Quantum Center, 100 Novaya Street, Skolkovo, Moscow region, 143025, Russia}

\date{\today}

\begin{abstract}

We analyze excitation of two ground state atoms to a double Rydberg state by a two-photon chirped optical pulse in the regime of adiabatic rapid passage. 
For intermediate Rydberg-Rydberg interaction 
strengths, relevant for atoms separated by $\sim$ten $\mu$m, adiabatic excitation can be achieved at experimentally feasible Rabi frequencies and chirp rates of the pulses, resulting in high transfer 
efficiencies. We also study the adiabatic transfer between ground and Rydberg states as a means to realize a controlled phase 
gate between atomic qubits. 

\end{abstract}

\maketitle

\section{Introduction}

Neutral atoms in ground hyperfine states represent a clean, well-controlled system for quantum information processing 
providing long time qubit storage, easy initialization, readout, manipulation with 
optical and magnetic fields, and scalability \cite{Neutral-atoms-review}. 
Two-qubit gates with atoms can be realized by transferring them into high-lying Rydberg states, in which the atoms can interact via strong and 
long-distance van der Waals and dipole-dipole interactions 
\cite{Saffman-Rydb-review}. 
In the seminal proposal \cite{Lukin-blockade}
of a controlled phase gate for atomic qubits based on interaction in Rydberg states two realizations have been discussed. The first one works for  
 close atoms in a regime when the interaction strength $V_{\rm int}$ exceeds the Rabi frequency $\Omega$ of excitation pulses coupling qubit states to the 
Rydberg ones: $V_{\rm int}>\Omega$. In this regime of a Rydberg blockade the strong interaction prevents excitation 
of a second atom if the first atom has been excited. 
The blockade has been experimentally demonstrated for two atoms in separate dipole traps \cite{Rydb-blockade-experim}, followed by realization of 
a CNOT gate \cite{CNOT-Rydb} and entanglement \cite{entanglement-Rydb}. A second approach to the controlled phase gate,
 discussed in \cite{Lukin-blockade}, applies to smaller interaction strengths $V_{\rm int} \lesssim \Omega$, 
for which the blockade cannot work. The gate then can be implemented by conditionally exciting both atoms to the Rydberg state, 
letting them interact to accumulate a $\pi$ phase shift, and deexciting back to their original qubit states. This allows to realize the gate 
between atoms separated by several sites in an optical lattice architecture and between atoms in distant individual microtraps, making the 
systems scalable. The second approach has not yet been realized experimentally and is currently a subject of active theoretical investigation 
\cite{Molmer-STIRAP-Rydb,Tommaso-opt-cont-CPHASE,Molmer}. Interactions in Rydberg states can also find applications in quantum simulation \cite{Hendrick-Rydb-quant-sim,Bloch}, 
quantum repeaters \cite{Quantum-repeators}, and in the realization of efficient and non-local nonlinearities \cite{Rydberg-nonlin}, 
down to a single photon level \cite{Gorshkov}.

\begin{figure}[h]
\center{
\includegraphics[width=2.3in]{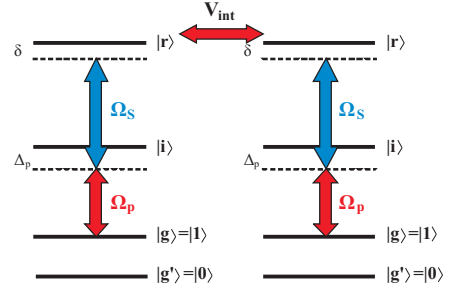}
\caption{\label{fig:level-scheme} Two ladder-type atomic systems interact with chirped pump (Rabi frequency $\Omega_{p}$) and Stokes fields 
(Rabi frequency $\Omega_{S}$) at $\ket{g}-\ket{i}$ and $\ket{i}-\ket{r}$ transitions, respectively. The one-photon and two-photon detunings are 
$\Delta_{p}=\omega_{ig}-\omega_{p}$ and 
$\delta=\omega_{rg}-\omega_{p}-\omega_{S}$. The atoms also experience a van der Waals or dipole-dipole 
interaction in the double Rydberg state $\ket{rr}$. Two long-lived ground states $\ket{g}$ and $\ket{g'}$ are used to encode qubit states $\ket{1}$ and $\ket{0}$.}
}
\end{figure}

Coherent control techniques such as STIRAP \cite{STIRAP-review} and adiabatic rapid passage (ARP) \cite{ARP-review} can provide robust excitation to the 
two-atom Rydberg state. Transfer of two atoms to the double Rydberg state by 
STIRAP  was studied in \cite{Molmer-STIRAP-Rydb}, where it was shown that in a 
system of three-level atoms, typically used in experiments, having ground $\ket{g}$, intermediate $\ket{i}$ and Rydberg states $\ket{r}$ (shown in 
Fig.\ref{fig:level-scheme}), the only dark state does not 
connect the two-atom ground $\ket{gg}$ and Rydberg $\ket{rr}$ states. Application of pump and Stokes pulses, resonant 
with the $\ket{g}-\ket{i}$ and $\ket{i}-\ket{r}$ transitions, respectively, transfers the system from $\ket{gg}$ to an
entangled $(\ket{ii}-\ket{gg})/\sqrt{2}$ state, not containing $\ket{rr}$. In a later work \cite{Molmer} it was found that 
a non-zero one-photon detuning $\Delta_{p} \sim V_{\rm int}$ in the STIRAP scheme produces a dressed state directly connecting $\ket{gg}$ to $\ket{rr}$, 
but the transfer efficiency to that state was not optimal because of population loss from a fast decaying intermediate state. 
STIRAP-like excitation to the $\ket{rr}$ state with larger detunings $\Delta_{p} \sim$ hundreds MHz can also be realized using optimal control 
techniques by shaping the pulses such that the population of the intermediate state is minimized \cite{Tommaso-opt-cont}.

Adiabatic rapid passage with chirped optical pulses is well-known for providing efficient population transfer between quantum states, and 
it will be studied in this work as a means to achieve robust excitation to double 
Rydberg states. 
In the two-photon ARP
excitation scheme of Fig.\ref{fig:level-scheme} a large one-photon detuning from the fast-decaying intermediate state can be used, allowing to minimize its population and obtain high 
transfer efficiencies. 
In section II we show that 
the two-atom system can be excited to the double Rydberg state along 
one of the dressed states, directly connecting $\ket{gg}$ to $\ket{rr}$. In section III we numerically calculate the transfer efficiency 
for Rb atoms taking into account decays from the intermediate and Rydberg states and analyze its dependence on Rabi frequencies and chirp rates of the pulses. 
Finally, 
in section IV we calculate the fidelity of the controlled phase gate which can be realized by conditionally transferring the ground state qubits to the $\ket{rr}$ state and 
 back such that the qubit state $\ket{11}$ accumulates the $\pi$ phase shift, and conclude in Section V.

\section{Dressed states for two three-level atoms interacting with a two-photon chirped field}

\begin{figure*}[h]
\center{
\includegraphics[width=6.in]{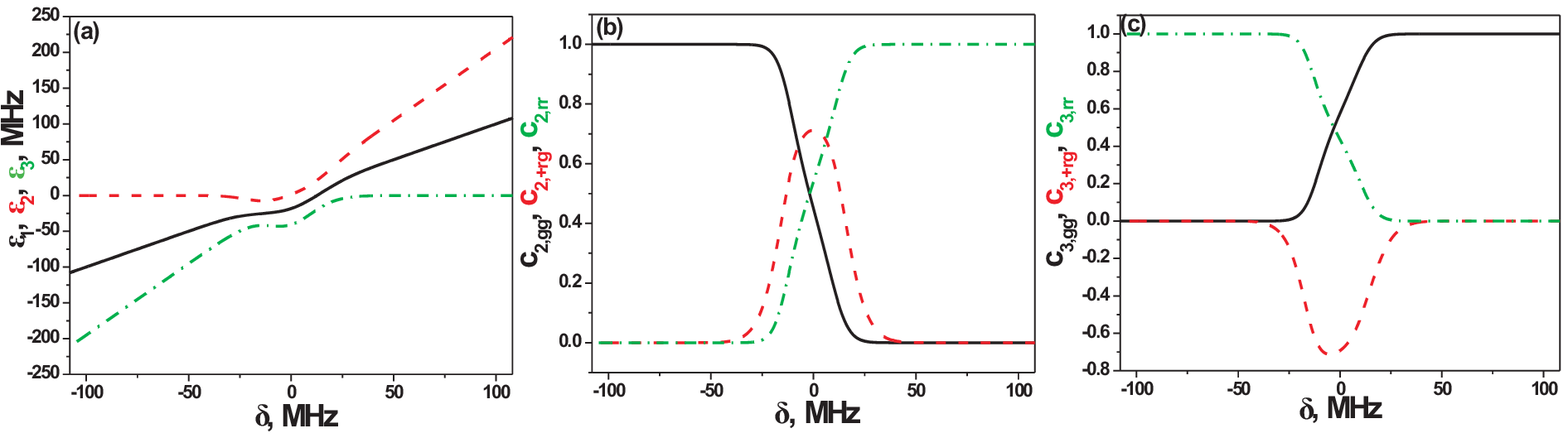}
\caption{\label{fig:dressed-states} (a) Energies $\epsilon_{1}$, $\epsilon_{2}$ and $\epsilon_{3}$ of the dressed states from Eq.(\ref{eq:eigenequation}),
 shown by solid black, dashed red and dash-dotted green lines, respectively; (b), (c) Components of the dressed states 
$\ket{\Psi_{2,3}}=c_{2,3,gg}\ket{gg}+c_{2,3,+rg}\ket{+,rg}+c_{2,3,rr}\ket{rr}$, where $c_{2,3,gg}$, $c_{2,3,+rg}$ and $c_{2,3,rr}$ are given by solid black, 
dashed red and dashed-dotted green curves, respectively. The Rabi frequencies and the detuning were parametrized as $\Omega_{p,S}=\Omega_{0p,0S}\exp(-(t-t_{c})^2/2\tau^{2}_{p,S})$ and 
$\delta=2\alpha(t-t_{c})$ with $\Omega_{0p}=\Omega_{0S}=120$ MHz, $\tau_{p}=\tau_{S}=75$ ns, $\alpha=190$ MHz/$\mu$s, $\Delta_{p}=1.5$ GHz, $V_{\rm int}=5$ MHz.}
}
\end{figure*}

We consider two three-level atoms with internal states $\ket{g}$, $\ket{i}$ and $\ket{r}$, shown in Fig.\ref{fig:level-scheme}. Each atom interacts with 
two chirped laser pulses, pump and Stokes, on the transitions $\ket{g}-\ket{i}$ and $\ket{i}-\ket{r}$, respectively.
In the Rydberg states $\ket{r}$ they additionally interact with each other via dipole-dipole or van der Waals interactions. The system can be 
described by the Schrodinger equation assuming that all the interactions are much faster compared to decays of intermediate and Rydberg states. 
For simplicity of the analysis one can use "molecular" states: $\ket{gg}$, 
$\ket{\pm}_{ig}=(\ket{ig}\pm \ket{gi})/\sqrt{2}$, $\ket{ii}$, $\ket{\pm}_{rg}=(\ket{rg}\pm \ket{gr})/\sqrt{2}$, 
$\ket{\pm}_{ri}=(\ket{ri}\pm \ket{ir})/\sqrt{2}$, $\ket{rr}$. Expanded in these states the two atom wave function has a form 
$\ket{\Psi}=c_{gg}\ket{gg}+\sum_{\sigma=\pm}c_{\sigma,ig}\ket{\sigma}_{ig}+c_{ii}\ket{ii}+\sum_{\sigma=\pm}c_{\sigma,rg}\ket{\sigma}_{rg}+\sum_{\sigma=\pm}c_{\sigma,ri}\ket{\sigma}_{ri}+c_{rr}\ket{rr}$, where the amplitudes 
of the "molecular" states are expressed via amplitudes of pure two atom states as follows: $c_{\pm,ig}=(c_{ig}\pm c_{gi})/\sqrt{2}$, 
$c_{\pm,rg}=(c_{rg}\pm c_{gr})/\sqrt{2}$, $c_{\pm,ri}=(c_{ri}\pm c_{ir})/\sqrt{2}$.

The Hamiltonian of the system in the rotating wave approximation is given by:
\begin{eqnarray}
\label{eq:Hamiltonian}
H/\hbar = \sum_{j=1,2}\left(\Delta_{p}(t)\left| i \right. \rangle _{j}\langle \left. i \right|+\delta(t)\left| r \right. \rangle _{j}\langle \left. r \right| \right)+V_{\rm int}\left| rr \right. \rangle \langle \left. rr \right| \nonumber \\
-\sum_{j=1,2}\left(\vec{\mu}_{ig}\vec{E}_{p}(t)/2\hbar\left| i \right. \rangle _{j}\langle \left. g \right|+\vec{\mu}_{ri}\vec{E}_{S}(t)/2\hbar\left| r \right. \rangle _{j}\langle \left. i \right|+{\rm H.c.}\right), 
\end{eqnarray}
where $\Delta_{p}(t)=\omega_{ig}-\omega_{p}(t)$, $\delta(t)=\omega_{rg}-\omega_{p}(t)-\omega_{S}(t)$ are the one and two-photon detunings of the 
pump and Stokes field frequencies $\omega_{p}$ and $\omega_{S}$ from the atomic transition
frequencies $\omega_{ig}$ and $\omega_{rg}$, $\vec{\mu}_{ig}$ and $\vec{\mu}_{ri}$ are the 
dipole moments of the corresponding transitions and $\vec{E}_{p,S}$ are the amplitudes of the pump and Stokes fields. 
The Schrodinger equations for the "molecular" state amplitudes are then given by:
\begin{eqnarray}
\label{eq:Schrod}
i\frac{dc_{gg}}{dt} & = & -\sqrt{2}\Omega_{p}c_{+,ig}, \nonumber \\
i\frac{dc_{+,ig}}{dt} & = & \Delta_{p}c_{+,ig}-\sqrt{2}\Omega_{p}c_{gg}-\sqrt{2}\Omega_{p}c_{ii}-\Omega_{S}c_{+,rg}, \nonumber \\
i\frac{dc_{-,ig}}{dt} & = & \Delta_{p}c_{-,ig}-\Omega_{S}c_{-,rg}, \nonumber \\
i\frac{dc_{ii}}{dt} & = & 2\Delta_{p}c_{ii}-\sqrt{2}\Omega_{p}c_{+,ig}-\sqrt{2}\Omega_{S}c_{+,ri}, \nonumber \\
i\frac{dc_{+,rg}}{dt} & = & \delta c_{+,rg}-\Omega_{S}c_{+,ig}-\Omega_{p}c_{+,ri}, \nonumber \\
i\frac{dc_{-,rg}}{dt} & = & \delta c_{-,rg}-\Omega_{S}c_{-,ig}-\Omega_{p}c_{-,ri}, \nonumber \\
i\frac{dc_{+,ri}}{dt} & = & (\delta+\Delta_{p})c_{+,ri}-\sqrt{2}\Omega_{S}c_{ii}-\sqrt{2}\Omega_{S}c_{rr}-\Omega_{p}c_{+,rg}, \nonumber \\
i\frac{dc_{-,ri}}{dt} & = & (\delta+\Delta_{p})c_{-,ri}-\Omega_{p}c_{-,rg}, \nonumber \\
i\frac{dc_{rr}}{dt} & = & (2\delta+V_{\rm int})c_{rr}-\sqrt{2}\Omega_{S}c_{+,ri}, 
\end{eqnarray} 
where $\Omega_{p}=\vec{\mu}_{ig}\vec{E}_{p}/2\hbar$, $\Omega_{S}=\vec{\mu}_{ri}\vec{E}_{S}/2\hbar$ are the Rabi frequencies of the pump and Stokes fields. 
First, one can notice from Eqs.(\ref{eq:Schrod}) that 
the $\ket{+}$ and $\ket{-}$ states decouple, i.e. laser fields connect only states within these subsystems. Second, the $\ket{gg}$ state is 
laser coupled only to $\ket{+}$ states such that if initially the atoms are in the $\ket{gg}$ state, the $\ket{-}$ subsystem is never 
populated, and, therefore, can be discarded in the present model which neglects decays. Next, we assume that the one-photon detuning is 
large $|\Delta_{p}|\gg \delta,\;\Omega_{p},\;\Omega_{S}$ 
such that the amplitudes $c_{+,\;ig}$, $c_{+,\;ii}$ and $c_{+,\;ri}$ can be replaced by their steady-state solutions as follows:
\begin{eqnarray}
c_{+,\;ig}\approx \frac{\sqrt{2}\Omega_{p}c_{gg}+\Omega_{S}c_{+,\;rg}+\sqrt{2}\Omega_{p}c_{ii}}{\Delta_{p}}, \nonumber \\
c_{+,\;ri}\approx \frac{\sqrt{2}\Omega_{S}c_{ii}+\Omega_{p}c_{+,\;rg}+\sqrt{2}\Omega_{S}c_{rr}}{\Delta_{p}}, \nonumber \\
c_{ii} \approx \frac{\sqrt{2}\Omega_{p}c_{+,\;ig}+\sqrt{2}\Omega_{S}c_{+,\;ri}}{2\Delta_{p}}. \nonumber
\end{eqnarray}
Expressing further the $c_{+,\;ig}$ and $c_{+,\;ri}$ in terms of $c_{gg}$, $c_{+,\;rg}$ and $c_{rr}$
we eliminate the intermediate state $\ket{i}$ and reduce the system Eqs.(\ref{eq:Schrod}) to three equations for states $\ket{gg}$, $\ket{+}_{rg}$ 
and $\ket{rr}$:
\begin{eqnarray}
i\frac{dc_{gg}}{dt}=-\frac{2\Omega_{p}^{2}}{\Delta_{p}}c_{gg}-\sqrt{2}\Omega c_{+,\;rg}-2\frac{\Omega^{2}}{\Delta_{p}}c_{rr}, \nonumber \\
i\frac{dc_{+,\;rg}}{dt}=\left(\delta-\frac{\Omega_{p}^{2}+\Omega_{S}^{2}}{\Delta_{p}}\right)c_{+,\;rg}-\sqrt{2}\Omega c_{gg}-\sqrt{2}\Omega c_{rr}, \nonumber \\
i\frac{dc_{rr}}{dt}=\left(\delta+V_{\rm int}-\frac{2\Omega_{S}^{2}}{\Delta_{p}}\right)c_{rr}-\sqrt{2}\Omega c_{+,\;rg}-\frac{2\Omega^{2}}{\Delta_{p}}c_{gg}, \nonumber
\end{eqnarray}
where $\Omega=\Omega_{p}\Omega_{S}/\Delta_{p}$ is the two-photon Rabi frequency. One can now obtain the dressed states of the two atom system and their 
energies $\epsilon$ from the energy equation:
\begin{align}
\label{eq:eigenequation}
-\left(\epsilon+\frac{2\Omega_{p}^{2}}{\Delta_{p}}\right)\left(\delta-\frac{\Omega_{p}^{2}+\Omega_{S}^{2}}{\Delta_{p}}-\epsilon\right)\left(2\delta-\frac{2\Omega_{S}^{2}}{\Delta_{p}}-\epsilon+V_{\rm int}\right)+ \nonumber \\
+2\Omega^{2}\left(2\epsilon+2\frac{\Omega_{p}^{2}+\Omega_{S}^{2}}{\Delta_{p}}-\frac{4\Omega^{2}}{\Delta_{p}}-2\delta-V_{\rm int}\right)=0. 
\end{align}
Neglecting the $\Omega^{2}/\Delta_{p}$ term and setting $V_{\rm int}=0$ allows one to obtain analytical expressions for the dressed states and their 
energies in the absence of the Rydberg-Rydberg interaction: 
\begin{eqnarray}
\label{eq:eigenenergies}
\epsilon_{1}=\delta -\frac{\Omega_{p}^{2}+\Omega_{S}^{2}}{\Delta_{p}}, \nonumber \\
\epsilon_{2,3}=\delta-\frac{\Omega_{p}^{2}+\Omega_{S}^{2}}{\Delta_{p}}\pm \sqrt{\left(\delta-\frac{\Omega_{p}^{2}+\Omega_{S}^{2}}{\Delta_{p}}\right)^{2}+4\delta\frac{\Omega_{p}^{2}}{\Delta_{p}}}= \nonumber \\
=\delta-\frac{\Omega_{p}^{2}+\Omega_{S}^{2}}{\Delta_{p}}\pm \sqrt{\delta^{2}+2\delta \frac{\Omega_{p}^{2}-\Omega_{S}^{2}}{\Delta_{p}}+\left(\frac{\Omega_{p}^{2}+\Omega_{S}^{2}}{\Delta_{p}}\right)^{2}},
\end{eqnarray}
where $\epsilon_{2,3}$ is related to the $\pm$ signs, respectively.

The corresponding  dressed states are the following:
\begin{eqnarray}
\ket{\Psi_{1}}=\frac{1}{\sqrt{\left(\delta+\frac{\Omega_{p}^{2}-\Omega_{S}^{2}}{\Delta_{p}}\right)^{2}+4\Omega^{2}}}\left(\sqrt{2}\Omega \ket{gg}- \right. \nonumber \\
\left. \left(\delta+\frac{\Omega_{p}^{2}-\Omega_{S}^{2}}{\Delta_{p}}\right)\ket{+}_{rg}-\sqrt{2}\Omega\ket{rr}\right), \nonumber \\
\ket{\Psi_{2,3}}=\frac{1}{2\sqrt{\left(\delta+\frac{\Omega_{p}^{2}-\Omega_{S}^{2}}{\Delta_{p}}\right)^{2}+4\Omega^{2}}}\left(\left(\epsilon_{3,2}+\frac{2\Omega_{p}^{2}}{\Delta_{p}}\right)\ket{gg} \right. \nonumber \\
\left. -2\sqrt{2}\Omega \ket{+}_{rg}-\left(\epsilon_{2,3}+\frac{2\Omega_{p}^{2}}{\Delta_{p}}\right)\ket{rr}\right). \nonumber \\
\end{eqnarray}
We can analyze the  dressed states in the limits of a large two-photon detuning $\delta$: $|\delta| \gg \Omega_{p}^{2}/\Delta_{p},\; \Omega_{S}^{2}/\Delta_{p}$:
\begin{eqnarray}
\label{eq:limits}
1)\;\; \delta>0: \nonumber \\
\ket{\Psi_{1}}\approx -\ket{+}_{rg},\;\; \ket{\Psi_{2}}\approx -\ket{rr}, \;\; \ket{\Psi_{3}} \approx \ket{gg}, \nonumber \\ 
2)\;\; \delta<0: \nonumber \\
\ket{\Psi_{1}}\approx \ket{+}_{rg}, \;\; \ket{\Psi_{2}} \approx -\ket{gg}, \;\; \ket{\Psi_{3}} \approx \ket{rr},
\end{eqnarray}
which shows that $\ket{\Psi_{3}}\approx \ket{gg}$ for large $\delta>0$ and $\ket{\Psi_{3}}\approx \ket{rr}$ 
for large $\delta<0$. As a result, one can transfer the two atoms from the $\ket{gg}$ state to 
the $\ket{rr}$ state using a negative chirp $d\delta/dt<0$ and back from the $\ket{rr}$ to the $\ket{gg}$ state using a positive chirp $d\delta/dt>0$. 
The same can be done using the $\ket{\Psi_{2}}$ state and a positive chirp to realize the $\ket{gg}\rightarrow \ket{rr}$ transfer, and a negative chirp to 
bring the system back into $\ket{gg}$.

Above it was assumed that the interaction in the Rydberg states is zero, which allowed us to obtain analytical expressions for dressed states and their energies. 
When $V_{\rm int}\ne 0$ the energies and dressed states can be calculated only numerically. We are interested in the case $V_{\rm int} \sim \Omega$, 
when the dipole blockade is not working, and investigate how two atoms can be transferred to the double Rydberg state in this regime. 
The energies and the amplitudes of the $\ket{gg}$, $\ket{+,rg}$ and $\ket{rr}$ components of the $\ket{\Psi_{2}}$ and $\ket{\Psi_{3}}$ states 
in the case $V_{\rm int}=0.5\Omega$ are shown in Fig.\ref{fig:dressed-states}a,b and c, respectively. 
One can see that for a large positive $\delta$ $\ket{\Psi_{3}}\approx \ket{gg}$ and for a large negative $\delta$ $\ket{\Psi_{3}}\approx \ket{rr}$ 
(for a large negative $\delta$ $\ket{\Psi_{2}}\approx \ket{gg}$ and for a large positive $\delta$ $\ket{\Psi_{2}}\approx \ket{rr}$),
as expected from Eq.(\ref{eq:limits}). Fig.\ref{fig:dressed-states} shows, therefore, that for intermediate interaction strengths the system can still be transferred from $\ket{gg}$ to $\ket{rr}$ if it adiabatically follows either $\ket{\Psi_{3}}$ for 
a negative chirp rate or $\ket{\Psi_{2}}$ for a positive one.

\section{Efficiency of excitation to double Rydberg state by ARP}

In this section we numerically analyze the efficiency of two-photon excitation from $\ket{gg}$ to a double Rydberg state $\ket{rr}$ using adiabatic rapid passage. 
We consider two three-level atoms 
interacting with pump and Stokes optical pulses and with each other via vdW or dipole-dipole interaction according to the 
Hamiltonian (\ref{eq:Hamiltonian}). 
We also take into account radiative decays from the intermediate $\ket{i}$ and Rydberg $\ket{r}$ states and describe the system using a density 
matrix equation:
\begin{eqnarray}
\label{eq:dens-matr}
\frac{d \rho}{dt}=\frac{i}{\hbar}\left[\rho,H\right]+{\cal L}\rho,
\end{eqnarray}
where the Lindblad term, incorporating the decays, is as follows:
\begin{eqnarray}
\label{eq:Lindblad}
{\cal L}\rho=\sum_{j=1,2}\frac{\Gamma_{i}}{2}\left(2\sigma_{ig}^{-\;j}\rho \sigma_{ig}^{+\;j}-\sigma_{ig}^{+\;j}\sigma_{ig}^{-\;j}\rho-\rho \sigma_{ig}^{+\;j}\sigma_{ig}^{-\;j}\right)+ \nonumber \\
+ \sum_{j=1,2}\frac{\Gamma_{r}}{2}\left(2\sigma_{ri}^{-\;j}\rho \sigma_{ri}^{+\;j}-\sigma_{ri}^{+\;j}\sigma_{ri}^{-\;j}\rho-\rho \sigma_{ri}^{+\;j}\sigma_{ri}^{-\;j}\right),
\end{eqnarray}
where $\sigma_{ig}^{+\;j}=\left|i\right. \rangle _{j}\langle \left. g \right|$, $\sigma_{ri}^{+\;j}=\left| r \right. \rangle _{j}\langle \left. i \right|$ and $\sigma_{ig}^{-\;j}=\left(\sigma_{ig}^{+\;j}\right)^{\dagger}$, 
$\sigma_{ri}^{-\;j}=\left(\sigma_{ri}^{+\;j}\right)^{\dagger}$ are the raising and lowering operators for the j$^{th}$ atom, $\Gamma_{i}$ and $\Gamma_{r}$ are 
radiative decay rates of the intermediate and Rydberg states. The pump and Stokes pulses Rabi frequencies have a Gaussian form $\Omega_{p,S}(t)=\Omega_{0p,0S}\exp(-(t-t_{c})^{2}/2\tau^{2}_{p,S})$ 
and detunings are $\Delta_{p}(t)=\omega_{ig}-\omega_{p}(t)=\Delta_{0p}+\alpha(t-t_{c})$, $\Delta_{S}(t)=\omega_{ri}-\omega_{S}(t)=\Delta_{0S}+\alpha(t-t_{c})$, 
where $\alpha$ is the linear chirp rate of the pulses, assumed equal for both.

\begin{figure}[h]
\center{
\includegraphics[width=3.5in]{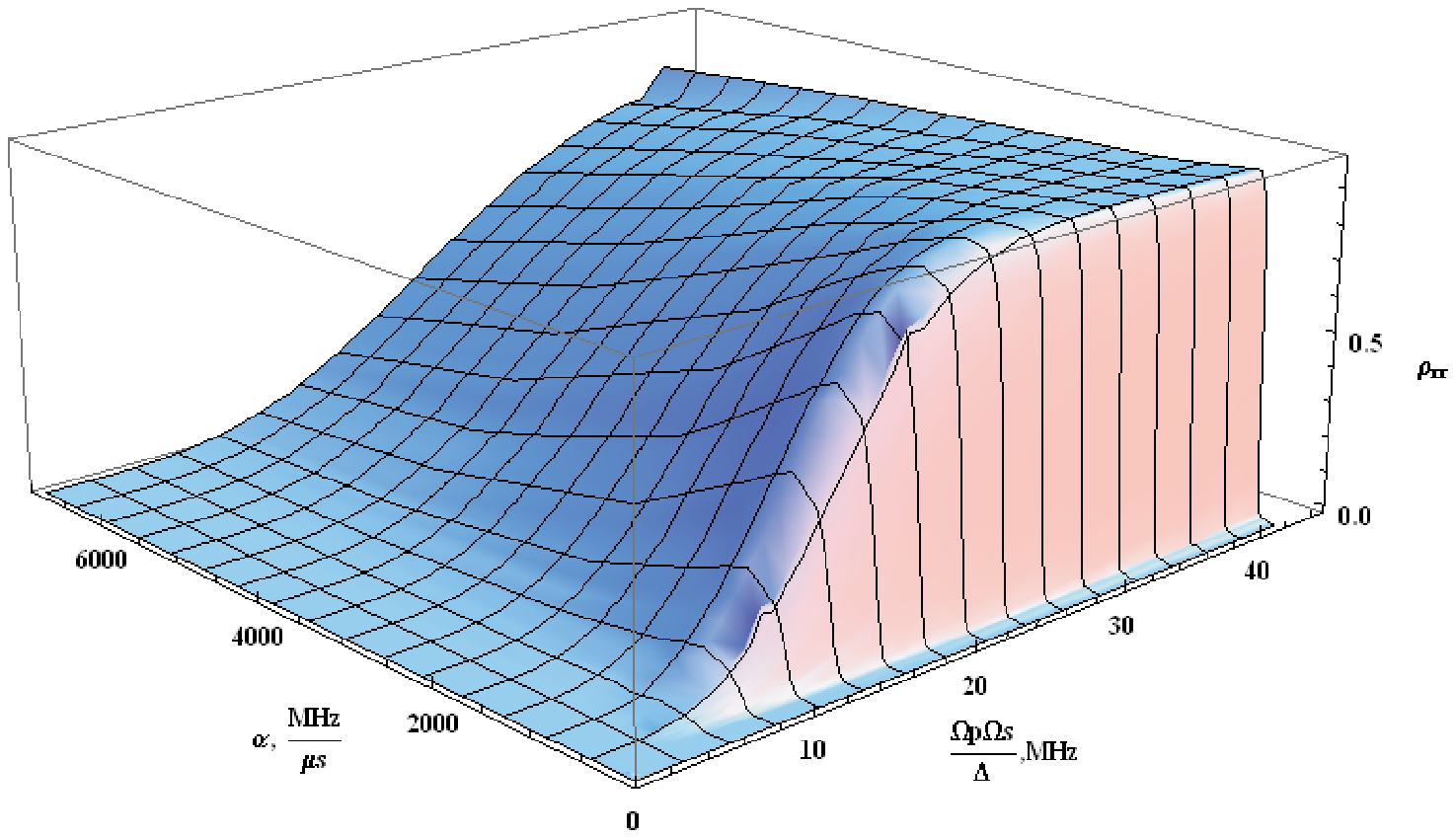}
\caption{\label{fig:transfer-eff} Transfer efficiency to the double Rydberg state, given by the population $\rho_{rr}$, as a function of the 
two-photon Rabi frequency and chirp rate. In the calculations $^{87}$Rb atoms were assumed with $\ket{i}$=5P$_{3/2}$ with decay rate 
$\Gamma_{i}=6$ MHz and $\ket{r}=80$S with $\Gamma_{r}=485$ Hz. Pump and Stokes pulses had Rabi frequencies 
$\Omega_{p,S}=\Omega_{0p,0S}\exp(-(t-t_{c})^2/2\tau^{2}_{p,S})$ with $\Omega_{0p}=\Omega_{0S}=250$ MHz, $\tau_{p}=\tau_{S}=100$ ns, 
$t_{c}=20/\Gamma_{i}$ and detunings $\Delta_{p,S}=\Delta_{0p,0S}+\alpha(t-t_{c})$ with $\alpha=475$ MHz/$\mu$s.  
Other parameters were the following: $V_{\rm int}=50$ MHz, $\Delta_{0p}=2.19$ GHz, $\Delta_{0S}=-2.268$ GHz.}
}
\end{figure}

\begin{figure}[h]
\center{
\includegraphics[width=3.in]{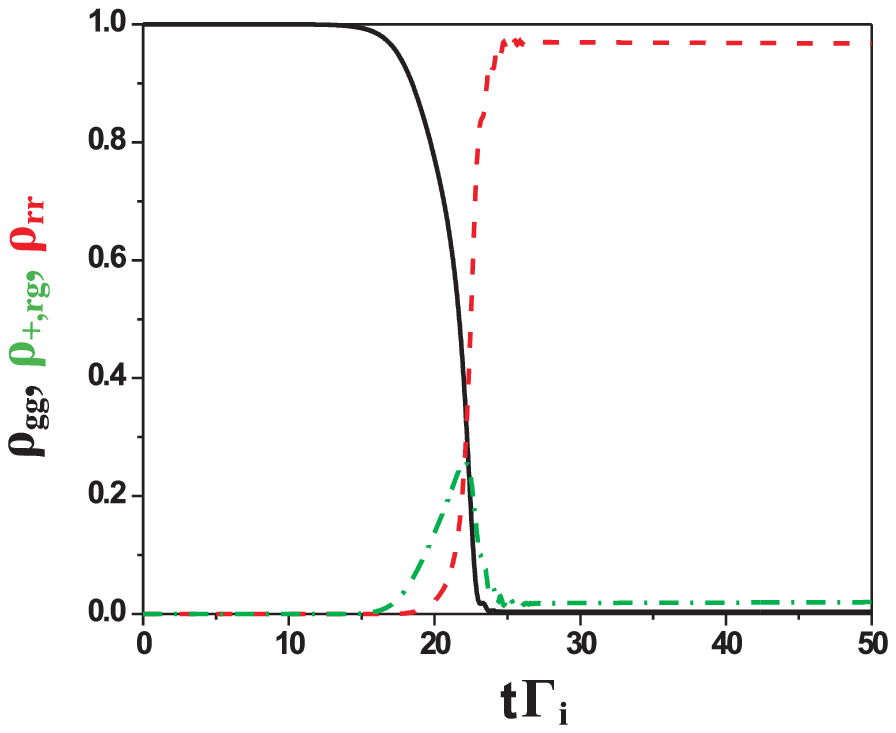}
\caption{\label{fig:transfer-time-dep} Time dependence of populations of the states $\ket{gg}$ (solid black curve), $\ket{+}_{rg}$ (green dash-dotted curve) and 
$\ket{rr}$ (red dashed curve) during transfer. In the calculations the same parameters of $^{87}$Rb atoms were assumed as in Fig.\ref{fig:transfer-eff}.}
}
\end{figure}

The population $\rho_{rr}$ of the double Rydberg state, i.e. the excitation efficiency, is shown in Fig.\ref{fig:transfer-eff} for a range of two-photon Rabi 
frequencies and chirp rates. In calculations parameters of $^{87}$Rb atoms were used with $\ket{i}$=5P$_{3/2}$ with decay rate 
$\Gamma_{i}=6$ MHz and $\ket{r}=80$S with $\Gamma_{r}=485$ Hz, which included the decay due to spontaneous emission ($\sim 300$ Hz) and due to interaction with 
black-body radiation at $T=300$ K ($\sim 185$ Hz). One can see that the efficiency reaches $\sim 97\%$ for sufficiently high two-photon Rabi frequencies and chirp rates, 
providing adiabatic interaction between atoms and laser pulses. Adiabaticity requires that  
$|d\delta/dt| \ll \Omega^{2}$ and $|d\delta/dt|\tau_{p,S}^{2} \gg 1$ \cite{Malinovsky}, where $d\delta/dt=2\alpha$ is the two-photon chirp rate, as well as equal 
Rabi frequencies $\Omega_{p}=\Omega_{S}$, pulse durations $\tau_{p}=\tau_{S}$ and chirp rates of the pump and Stokes pulses. 
 From Fig.\ref{fig:transfer-eff}a one can see that the efficiency becomes high ($\sim 90\%$) 
for $\alpha \ge 300$ MHz/$\mu$s and $\Omega \ge 30$ MHz because for these parameters 
(using pulse durations $\tau_{p}=\tau_{S} \sim 0.1$ $\mu$s)  the adiabaticity 
conditions are satisfied: $\Omega^{2}/2\alpha \approx 4.5$ and $2\alpha \tau^{2}_{p,S}\approx 38$. 
Fig.\ref{fig:transfer-time-dep} gives the time evolution of populations of the $\ket{gg}$, $\ket{+}_{rg}$ and $\ket{rr}$ states during the excitation. 
The parameters of the pulses providing efficient transfer are challenging but within experimental reach: currently the Rabi frequencies of the pump and Stokes pulses 
can be increased up to $\Omega_{0p,0S}\sim 250$ MHz \cite{High-Rabi}, 
while the chirp rates can be as high as $\sim 40$ GHz/$\mu$s \cite{High-chirp-rates}.

\section{Controlled phase gate using ARP excitation to double Rydberg state}

The interaction in Rydberg states can be used to realize a controlled phase gate $C_{Z}$, which acts on two-qubit states $\ket{nm}$ ($n,m=0,1$) as 
$\ket{nm}\rightarrow e^{i\pi nm}\ket{nm}$. The gate can be implemented using ARP excitation in the following way. In the first step four-level atoms, 
shown in Fig.\ref{fig:level-scheme}, interact with a chirped two-photon 
pulse such that the system initially in $\ket{gg}=\ket{11}$ evolves along the $\ket{\Psi_{3}}$ 
dressed state. At the end of the pulse the atoms will be transferred into $\ket{rr}$ and acquire a phase factor $\int \epsilon_{3}(t)dt$. 
At the same time the states $\ket{gg'}=\ket{10}$ and $\ket{g'g}=\ket{01}$ will evolve into $\ket{r0}$ and $\ket{0r}$, respectively, and acquire a phase factor 
$\int \epsilon_{-}(t)dt$, where 
$\epsilon_{\pm}=\delta/2-\left(\Omega_{p}^{2}+\Omega_{S}^{2}\right)/2\Delta_{p} \pm \sqrt{\left(\delta/2\right)^{2}+\Omega^{2}}$ are the dressed state energies 
for a single atom interacting with the two-photon chirped pulse \cite{Our-PhysScript}.  
One can see from Eq.(\ref{eq:eigenenergies}) that if $V_{\rm int}=0$ and $\Omega_{p}=\Omega_{S}$, $\epsilon_{2,3}=2\epsilon_{\pm}$ as expected. When 
$V_{\rm int}\neq 0$ we can estimate the eigenenergies using Eq.(\ref{eq:eigenequation}) in the limit of small interaction strengths assuming 
$V_{\rm int} << |\delta|,\; \Omega_{p}^{2}/|\Delta_{p}|, \;\Omega_{S}^{2}/|\Delta_{p}|$: 
\begin{eqnarray}
\label{eq:pi-phase}
\epsilon_{2,3}=\delta-\frac{2\Omega_{p}^{2}}{\Delta_{p}} \pm \sqrt{\left(\delta+\frac{V_{\rm int}}{2}\right)^{2}+4\Omega^{2}} \approx  \nonumber \\
\approx \delta-\frac{2\Omega_{p}^{2}}{\Delta_{p}} \pm \sqrt{\delta^{2}+4\Omega^{2}} \pm \frac{\delta V_{\rm int}}{2\sqrt{\delta^{2}+4\Omega^{2}}}= \nonumber \\
=2\epsilon_{\pm} \pm \frac{\delta V_{\rm int}}{2\sqrt{\delta^{2}+4\Omega^{2}}},
\end{eqnarray}
where $\Omega_{p}=\Omega_{S}$ was set again. It shows that for $V_{\rm int} \ne 0$, the phase accumulated by the state $\ket{11}$, when it is transferred 
to $\ket{rr}$, differs from twice the phase of the states $\ket{01}$ and $\ket{10}$ when they are transferred to $\ket{0r}$ and $\ket{r0}$, respectively. 
We also assume that the state $\ket{00}$ is not interacting with the pulse, which can be 
realized e.g. by choosing a specific polarization of the pump field. As a result, the state $\ket{00}$ will acquire zero phase.

\begin{figure}[h]
\center{
\includegraphics[width=2.5in]{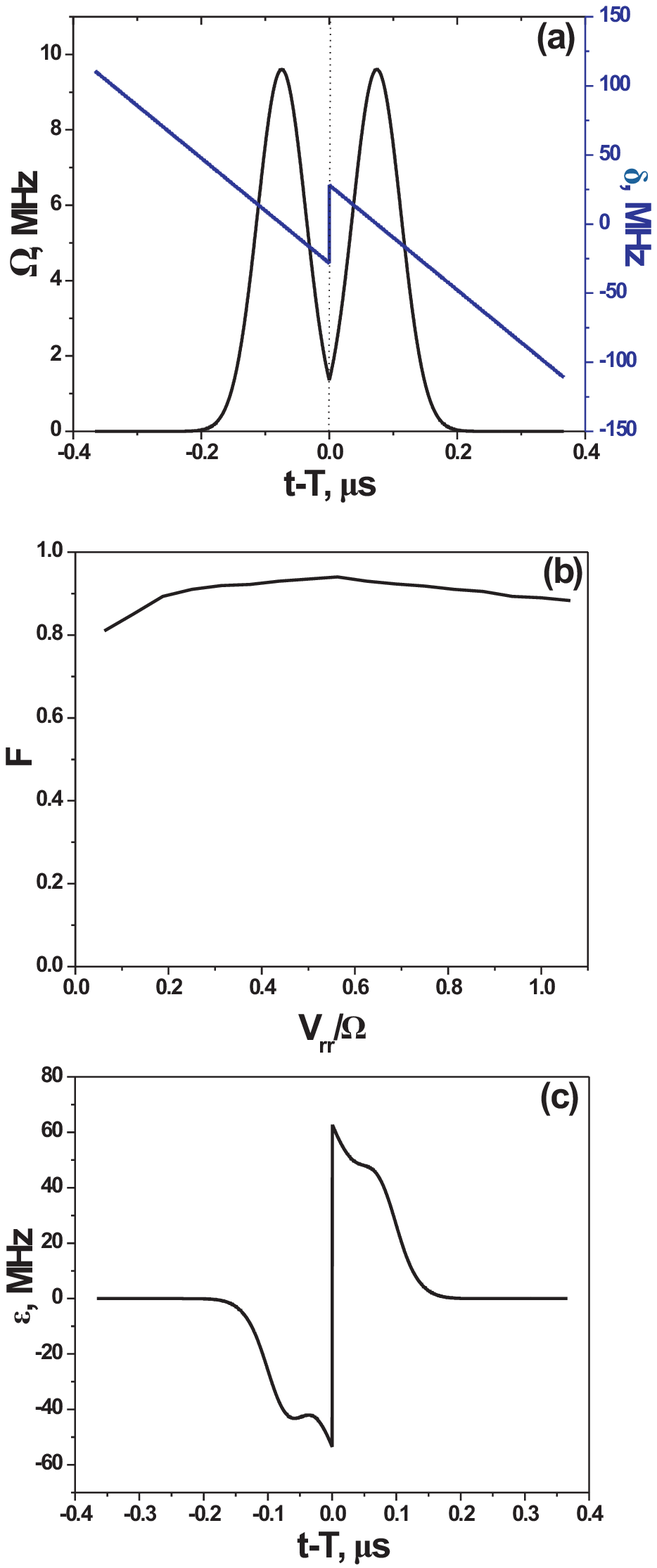}
\caption{\label{fig:fidelity} (a) Time dependence of the two-photon Rabi frequency and chirp rate during the controlled phase gate for $V_{\rm int}=5$ MHz; 
 (b) Fidelity of the controlled phase gate. 
In the calculations $^{87}$Rb atoms were assumed with $\ket{i}$=5P$_{3/2}$ with decay rate 
$\Gamma_{i}=6$ MHz and $\ket{r}=80$S with $\Gamma_{r}=485$ Hz. Parameters of the pulses: 
$\Omega_{0p}=\Omega_{0S}=120$ MHz, $\tau_{p}=\tau_{S}=75$ ns, $\alpha=190$ MHz/$\mu$s, $\Delta_{p}=1.5$ GHz; 
(c) Time dependence of the dressed state energy during the gate for the same interaction strength. The integral $\int_{T-\tau_{\rm st}}^{T+\tau_{\rm st}}\epsilon(t)dt$ gives the 
phase $\phi_{11}=\pi$.}
}
\end{figure}

In the second step the system returns back into qubit subspace with useful phases. To realize it one can apply the following trick \cite{Ryabtsev-ARP}: 
(i) the chirp in the first and the second 
steps has the same sign such that $\delta(T+t)=-\delta(T-t)$, where $T$ is the time boundary between the steps, as shown in Fig.\ref{fig:fidelity}a.  
The system then returns from $\ket{rr}$ to $\ket{gg}$ along $\ket{\Psi_{2}}$, and 
from $\ket{r0}$ and $\ket{0r}$ to $\ket{10}$ and $\ket{01}$ 
along $\epsilon_{+}$; (ii) the one-photon detuning changes sign in the second step with respect to the first one $\Delta_{p}\rightarrow -\Delta_{p}$. 
Provided the pump and Stokes pulses are applied symmetrically in time $\Omega_{p,S}(T+t)=\Omega_{p,S}(T-t)$ 
(see Fig.\ref{fig:fidelity}a), the conditions (i) and (ii) 
allow to cancel the overall phase accumulated by the $\ket{10}$ and $\ket{01}$ states: $\phi_{01}=\phi_{10}=\int_{\rm step I} \epsilon_{-}(t)dt+\int_{\rm step II}\epsilon_{+}(t)dt=0$. At the 
same time, the phase accumulated by the $\ket{11}$ state $\phi_{11}=\int_{\rm step I} \epsilon_{3}(t)dt+\int_{\rm step II}\epsilon_{2}dt \ne 0$. By adjusting the pulse 
parameters and interpulse time $\phi_{11}=\pi(2n+1)$ can be realized, which will produce the gate. For small interaction strengths one can estimate the 
$\phi_{11}$ phase using Eq.(\ref{eq:pi-phase}):
\begin{eqnarray}
\phi_{11} & = & \int_{T-\tau_{\rm st}}^{T}\epsilon_{3}(t)dt+\int_{T}^{T+\tau_{\rm st}}\epsilon_{2}(t)dt= \nonumber \\
&& =\int_{T-\tau_{\rm st}}^{T}\left(2\epsilon_{-}(t)-\frac{\delta V_{\rm int}}{2\sqrt{\delta^{2}+4\Omega^2}}\right)dt+ \nonumber \\
&& +\int_{T}^{T+\tau_{\rm st}}\left(2\epsilon_{+}(t)+\frac{\delta V_{\rm int}}{2\sqrt{\delta^{2}+4\Omega^2}}\right)dt= \nonumber \\
&& =-\int_{T-\tau_{\rm st}}^{T}\frac{\delta V_{\rm int}}{2\sqrt{\delta^{2}+4\Omega^2}}dt+\int_{T}^{T+\tau_{\rm st}}\frac{\delta V_{\rm int}}{2\sqrt{\delta^{2}+4\Omega^2}}dt= \nonumber \\
&& =2\int_{T}^{T+\tau_{\rm st}}\frac{\delta V_{\rm int}}{2\sqrt{\delta^{2}+4\Omega^2}}dt,
\end{eqnarray}
where $\tau_{\rm sp}$ is the time duration of each step, assumed equal, and $\delta(T+t)=-\delta(T-t)$ was applied in the last line. 
Using linearly chirped pulses with $\delta=2\alpha(t-t_{c})$ where $T<t_{c}<T+\tau_{\rm st}$ and setting $\Omega={\rm const}$ for simplicity we have
\begin{eqnarray} 
\phi_{11}=\frac{V_{\rm int}}{\alpha}\sqrt{\delta^{2}+4\Omega^{2}}\left|_{T}^{T+\tau_{\rm st}}= \right. \nonumber \\
=\frac{V_{\rm int}}{\alpha}\left(\sqrt{4\alpha^{2}(T+\tau_{\rm st}-t_{c})^{2}+4\Omega^{2}}-\sqrt{4\alpha^{2}(T-t_{c})^{2}+4\Omega^{2}}\right). \nonumber
\end{eqnarray}
In the limit $\alpha|T-t_{c}| \gg \Omega$  one obtains $\phi_{11}\approx 2V_{\rm int}\tau_{\rm st}$, showing that the duration of each step is $\tau_{\rm st}=\pi/2V_{\rm int}$, 
of the same order as the time between two STIRAP pulse sequencies required to accumulate a $\pi$ phase shift in \cite{Molmer}. For small 
interaction strengths the gate duration can become comparable to the Rydberg state decay time, which will reduce gate fidelity. High fidelities are expected 
for intermediate strengths $V_{\rm int} \sim \Omega$, and Fig.\ref{fig:fidelity}b shows the fidelity 
$F=\langle \left.\Psi_{\rm ideal} \right| \rho \left(T+\tau_{\rm st}\right) \left|\Psi_{\rm ideal} \right. \rangle$ numerically calculated in this regime, 
where the expected state of the system in the absence of errors is $\ket{\Psi_{\rm ideal}}=\frac{1}{2}\left(\ket{00}+\ket{01}+\ket{10}-\ket{11}\right)$ 
and $\rho \left(T+\tau_{\rm st} \right)$ is the density matrix of the system at the end of the second step taking into account population and coherence decays. 
An 
initial state of the system $\ket{\Psi(T-\tau_{\rm st})}=\frac{1}{2}\left(\ket{0}+\ket{1}\right) \bigotimes \left(\ket{0}+\ket{1}\right)=\frac{1}{2}\left(\ket{00}+\ket{01}+\ket{10}+\ket{11}\right)$ 
was used and the density matrix evolution was modelled by Eq.(\ref{eq:dens-matr}) with decays given by the Lindblad term (\ref{eq:Lindblad}), in which 
we assumed for simplicity that the population in the Rydberg state decayed to the intermediate state and in the intermediate state to $\ket{g}=\ket{1}$, i.e. 
there was no decay into the $\ket{g'}=\ket{0}$ state. It was also assumed that the $\ket{0}$ state was not 
interacting with the chirped pulse. The same parameters of $^{87}$Rb were used: the radiative decay rates $\Gamma_{i}=6$ MHz and 
$\Gamma_{r}=485$ Hz, corresponding to the 5P$_{3/2}$ and 80S states \cite{Rydberg-decay-rates}, respectively, and the energy splitting between the qubit 
states of $6.835$ GHz, corresponding to the hyperfine splitting between ground state $F=2$ and $F=1$ sublevels. During calculations the conditions of the first and second 
steps were applied: the Rabi frequencies were symmetric, and chirp rate and one-photon detuning antisymmetric with respect to the step time boundary. In order to 
accumulate the $\pi$ phase shift in the $\ket{11}$ state the time delay between two pulse sequencies $2(t_{c}-T)$ was adjusted every time $V_{\rm int}$ 
was changed. Fig.\ref{fig:fidelity}c shows time dependence of the dressed state energy during both steps for $V_{\rm int}=0.5\Omega$, with the 
phase $\phi_{11}$ given by the integral 
$\int_{T-\tau_{\rm st}}^{T+\tau_{\rm st}}\epsilon(t) dt$. 
Fidelities $\sim 94\%$ were obtained for $V_{\rm int}/\Omega \sim 0.5$, limited by an incomplete conversion of $\ket{gg}$ into $\ket{rr}$ during the first step, 
i.e. less than $100\%$ transfer efficiency $\rho_{rr}$,  
which resulted in a small admixture of the $\ket{\Psi_{3}}$ dressed state to the $\ket{\Psi_{2}}$ state during the second step, when the system returned from 
$\ket{rr}$ to $\ket{gg}$. The obtained fidelities are comparable to the ones expected in the STIRAP-based excitation scheme \cite{Molmer}. 
We checked the effect of the intermediate state decay on the fidelity, which was an important source of error in  
\cite{Molmer}, and found that complete cancellation of the decay gives the fidelity increase $\sim 1\%$. The intermediate state decay 
is less important in our scheme due to a large one-photon detuning. Decay from the Rydberg state also does not significantly affect the gate in our case 
due to its short 
duration $\sim 2(t_{c}-T)\approx 300$ ns and small decay rate $\Gamma_{r}=300$ Hz. 

Another possible way to implement the controlled phase gate is to follow the same dressed state, e.g. the $\ket{\Psi_{3}}$, in both steps, which can be done if the 
chirp rate changes sign in the second step such that $\delta(T+t)=\delta(T-t)$. In this case at the end of the second step 
the state $\ket{11}$ will acquire the phase shift $\phi_{11}=\int_{T-\tau_{\rm st}}^{T+\tau_{\rm st}}\epsilon_{3}(t)dt$, and the $\ket{01}$ and $\ket{10}$ states will 
acquire the shift $\phi_{10}=\phi_{01}=\int_{T-\tau_{\rm st}}^{T+\tau_{\rm st}}\epsilon_{-}(t)dt$. The simplest way to realize this is to adjust pulse parameters in 
such a way that $\phi_{01,10}=\pi$ and $\phi_{11}=3\pi$ (for small $V_{\rm int}$ $\phi_{11}\approx 2\phi_{10,01}$, for larger $V_{\rm int}$ $\phi_{11}>2\phi_{01,10}$).  
However, this scheme might be more challenging than the one discussed above, because two phases have to be simultaneously tuned to specific values.

The above fidelity calculations assume that the atomic motion is frozen during the gate. This assumption can be violated due to mechanical forces acting on atoms 
in the double Rydberg state \cite{Lesanovsky-forces}. The forces can result in excitation of higher motional states for atoms trapped in microtraps and optical lattices, 
resulting in undesirable entanglement between the motional and qubit states. We estimate 
the probability of excitation from a ground to the first excited motional state for atoms in an optical lattice. The amplitude of the first motional state 
after the system is deexcited from the double Rydberg state is $c_{\rm mot}\sim {\cal F}\delta r(1-\exp(-2i\omega_{0}(t_{c}-T)))/\omega_{0}$, 
where ${\cal F}=\partial V_{\rm int}/\partial r=6C_{6}/r^{7}=6V_{\rm int}/r$, is the force acting between two atoms interacting via the van der Waals $C_{6}/r^{6}$ 
interaction, $\omega_{0}$ is the trapping potential oscillation frequency and $\delta r \sim \sqrt{\hbar/m\omega_{0}}$ is the ground motional state wavefunction 
width, $r$ is the distance between atoms, and $2(t_{c}-T)$ is the time the system spends in the double Rydberg state. Assuming $V_{\rm int}=5$ MHz, corresponding to $r=9.3$ $\mu$m 
for atoms in the 80S state \cite{Robin-C6}, $\omega_{0}=100$ kHz, $\delta r=35$ nm and 
$t_{c}-T \approx 150$ ns, the probability of the motional state excitation $|c_{\rm mot}|^{2}\sim 0.02$, which gives the additional error in the fidelity.

Our analysis shows that the ARP and STIRAP-type excitation \cite{Molmer} to the double Rydberg state, which use simple analytic pulse sequencies, 
predict similar high controlled phase gate fidelities 
$\sim 94\%$ and $\sim 97\%$ in the former and the latter cases, respectively. However, these values are not good enough to allow fault-tolerant quantum 
computation, which requires the gate fidelity $> 99.9\%$ \cite{Fault-tolerant-QC}. One of the strategies to increase the fidelity is to 
apply more complex coherent control techniques such as optimal control \cite{OCT} and genetic \cite{GA} algorithms to shape laser pulses. Optimal control of 
STIRAP based blockaded controlled phase gate has been analyzed recently in \cite{High-Rabi}, where it was shown that the gate error can be decreased by an order of magnitude (from $10^{-3}$ to $10^{-4}$) 
if one uses optimized pulse sequences instead of analytic. Optimization of chirped pulses, first proposed in \cite{ARP-opt-cont}, 
is succesfully used to achieve efficient population transfer between molecular states \cite{ARP-optimized-schemes} and might help to improve the fidelity of the 
ARP based gate, which will be the subject of a future work.

\section{Conclusion}

In conclusion, we analyzed excitation of two ground state atoms to a double Rydberg state by a chirped two-photon pulse using adiabatic rapid passage. During ARP-type  
excitation dressed 
states of the coupled atoms-light system provide direct connection of the $\ket{gg}$ 
to the double Rydberg $\ket{rr}$ state contrary to the 
case of resonant STIRAP \cite{Molmer-STIRAP-Rydb}. Numerical analysis taking into account population and coherence decays predicts robust transfer
 to the $\ket{rr}$ state that can reach high efficiency $\sim 97\%$ in the case of $^{87}$Rb atoms for 
intermediate interaction strengths $V_{\rm int}\sim \Omega$. 
The high transfer efficiency is possible due to a large one-photon 
detuning allowed in the ARP scheme, minimizing losses from the fast-decaying intermediate state. The large one-photon detuning has to be compensated by 
high Rabi frequencies $\sim 200$ MHz of the pump and Stokes pulses to achieve 
adiabaticity, but they are currently within experimental reach along with required chirp rates $\sim$hundreds MHz/$\mu$s. 
We also considered a controlled phase gate for two atomic qubits based on ARP transfer to interacting Rydberg states. Applying antisymmetric 
one- and two-photon detunings and symmetric Rabi frequencies during excitation and deexcitation steps one can cancel the phases of the $\ket{01}$ and 
$\ket{10}$ qubit states and tune the phase of the $\ket{11}$ state to $\pi$, producing the gate. Gate fidelities $\sim 94\%$ were numerically 
predicted at $V_{\rm int}\sim \Omega$ for $^{87}$Rb atoms, limited by incomplete switching of the dressed states between the excitation and deexcitation steps.  
Our analysis shows that ARP and STIRAP-type double Rydberg state excitations using simple analytic pulse sequencies are expected to achieve comparable transfer efficiencies and 
controlled phase gate fidelities, which are high but still insufficient for fault-tolerant quantum computation. One of the ways to increase the transfer 
efficiency and therefore 
the gate fidelity is to use more complex optimized chirped pulses.

\section{Acknowledgements} 

The author thanks Svetlana Malinovskaya for fruitful discussions, the Russian Quantum Center and the Russian Fund 
for Basic Research (grant RFBR 14-02-00174) for financial support.


\begin{thebibliography}{11}

\bibitem{Neutral-atoms-review} A. Negretti, P. Treutlein, T. Calarco, Quant. Inf. Process. {\bf 10}, 721 (2011).

\bibitem{Saffman-Rydb-review} M. Saffman, T. G. Walker, K. Molmer, Rev. Mod. Phys. {\bf 82}, 2313 (2010).

\bibitem{Lukin-blockade} D. Jaksch, J. I. Cirac, P. Zoller, S. L. Rolston, R. Cote, M. D. Lukin, Phys. Rev. Lett. {\bf 85}, 2208 (2000).

\bibitem{Rydb-blockade-experim} E. Urban, T. A. Jonhson, T. Henage, L. Isenhower, D. D. Yavuz, T. G. Walker, M. Saffman. Nat. Phys. {\bf 5}, 
110 (2009); A. Gaetan, Y. Miroshnychenko, T. Wilk, A. Chotia, M. Viteau, D. Comparat, P. Pillet, A. Browaeys, P. Grangier, Nat. Phys. {\bf 5}, 
115 (2009).

\bibitem{CNOT-Rydb} L. Isenhower, E. Urban, X. L. Zhang, A. T. Gill, T. Henage, T. A. Johnson, T. G. Walker, M. Saffman, Phys. Rev. Lett. {\bf 104}, 
010503 (2010).

\bibitem{entanglement-Rydb} T. Wilk, A. Gaetan, C. Evellin, J. Wolters, Y. Miroshnychenko, P. Grangier, A. Browaeys, Phys. Rev. Lett. {\bf 104}, 
010502 (2010). 

\bibitem{Molmer-STIRAP-Rydb} D. Moller, L. B. Madsen, K. Molmer, Phys. Rev. Lett. {\bf 100}, 170504 (2008). 

\bibitem{Tommaso-opt-cont-CPHASE} M. H. Goerz, T. Calarco, C. P. Koch, J. Phys. B {\bf 44}, 154011 (2011).

\bibitem{Molmer} D. D. Bhaktavatsala Rao, K. Molmer, Phys. Rev. A {\bf 89}, 030301 (2014).

\bibitem{Bloch} P. Schaub, M. Cheneau, M. Endres, T. Fukuhara, S. Hild, A. Omran, T. Pohl, C. Gross, S. Kuhr, I. Bloch, Nature {\bf 491}, 87 (2012).

\bibitem{Hendrick-Rydb-quant-sim} H. Weimer, M. Muller, I. Lesanovsky, P. Zoller, H. P. Buchler, Nat. Phys. {\bf 6}, 382 (2010).

\bibitem{Quantum-repeators} Y. Han, B. He, K. Heshami, C.-Z. Li, C. Simon, Phys. Rev. A {\bf 81}, 052311 (2010); 
B. Zhao, M. Muller, K. Hammerer, P. Zoller, Phys. Rev. A {\bf 81}, 052329 (2010).

\bibitem{Rydberg-nonlin} J. D. Pritchard, D. Maxwell, A. Gauguet, K. J. Weatherhill, M. P. A. Jones, C. S. Adams, Phys. Rev. Lett. {\bf 105}, 193603 (2010); 
S. Svencli, N. Henkel, C. Ates, T. Pohl, Phys. Rev. Lett. {\bf 107}, 153001 (2011). 

\bibitem{Gorshkov} O. Firstenberg, T. Peyronel, Q.-Y. Liang, A. V. Gorshkov, M. D. Lukin, V. Vuletic, Nature {\bf 502}, 71 (2013).


\bibitem{STIRAP-review} K. Bergmann, H. Theuer, B. W. Shore, Rev. Mod. Phys. {\bf 70}, 1003 (1998). 


\bibitem{ARP-review} N. V. Vitanov, T. Halfmann, B. W. Shore, K. Bergmann, Ann. Rev. Phys. Chem. {\bf 52}, 763 (2001). 

\bibitem{Tommaso-opt-cont} M. M. Muller, H. R. Haakh, T. Calarco, C. P. Koch, C. Henkel, Quant. Inf. Proc. {\bf 10}, 711 (2011).

\bibitem{Deutsch-Rydb-CPHASE} T. Keating, R. L. Cook, A. Hankin, Y.-Y. Jan, G. W. Biedermann, I. H. Deutsch arxiv:1411.2622.

\bibitem{Malinovsky} V. S. Malinovsky, J. L. Krause, Europ. Phys. J. D {\bf 14}, 147 (2001).

\bibitem{High-Rabi} M. H. Goerz, E. J. Halperin, J. M. Aytac, C. P. Koch, K. B. Whaley, Phys. Rev. A {\bf 90} 032329 (2014).

\bibitem{High-chirp-rates} C. E. Rogers III, M. J. Wright, J. L. Carini, J. A. Pechkis, P. L. Gould, J. Opt. Soc. Am. B {\bf 24}, 1249 (2007).

\bibitem{Our-PhysScript} E. Kuznetsova, G. Liu, S. Malinovskaya, Phys. Script. T {\bf 160}, 014024 (2014).

\bibitem{Ryabtsev-ARP} I. I. Beterov, M. Saffman, E. A. Yakshina, V. P. Zhukov, D. B. Tretyakov, V. M. Entin, I. I. Ryabtsev, C. W. Mansell, C. MacCormick, 
S. Bergamini, M. P. Fedoruk, Phys. Rev. A {\bf 88}, 010303 (2013).

\bibitem{Rydberg-decay-rates} V. D. Ovsiannikov, I. L. Glukhov, E. A. Nikipelov, J. Phys. B {\bf 44}, 195010 (2011).

\bibitem{Lesanovsky-forces} W. Li, C. Ates, I. Lesanovsky, Phys. Rev. Lett. {\bf 110}, 213005 (2013).

\bibitem{Robin-C6} K. Singer, J. Sanojevic, M. Weidemuller, R. Cote, J. Phys. B {\bf 38}, S295 (2005).

\bibitem{Fault-tolerant-QC} B. W. Reichardt, Algorithmica {\bf 55}, 517 (2009). 

\bibitem{OCT} P. von den Hoff, S. Thallmair, M. Kowalewski, R. Siemering, R. de Vivie-Riedle, Phys. Chem. Chem. Phys. {\bf 14}, 14460 (2012).

\bibitem{GA} R. S. Judson, H. Rabitz, Phys. Rev. Lett. {\bf 68}, 1500 (1992).

\bibitem{ARP-opt-cont} B. Amstrup, J. D. Doll, R. A. Sauerbrey, G. Szabo, A. Lorincz, Phys. Rev. A {\bf 48}, 3830 (1993).

\bibitem{ARP-optimized-schemes} C. J. Bardeen, V. V. Yakovlev, K. R. Wilson, S. D. Carpenter, P. M. Weber, W. S. Warren, Chem. Phys. Lett. {\bf 280}, 
151 (1997); T. Hornung, R. Meier, M. Motzkus, Chem. Phys. Lett. {\bf 326}, 445 (2000).



\end{thebibliography}
\end{document}